\documentclass[%
reprint,
 superscriptaddress,
showpacs,
 amsmath,amssymb,
 aps,
]{revtex4-1}

\usepackage{graphicx}
\usepackage{dcolumn}
\usepackage{bm}
\usepackage{color}
\usepackage[usenames,dvipsnames]{xcolor}

\usepackage[normalem]{ulem}

\begin{document}

\preprint{APS/123-QED}

\title{Ultrafast x-ray-induced nuclear dynamics in diatomic molecules using femtosecond x-ray/x-ray pump-probe spectroscopy}

\author{C. S. Lehmann}
\altaffiliation{Present address:  Fachbereich  Chemie,  Philipps-Universit\"at  Marburg,  Marburg,  Germany}
\author{A. Pic\'on}
\email{antonio.picon.alvarez@gmail.com}
\affiliation{Argonne National Laboratory, Argonne, Illinois 60439, USA}
\author{C. Bostedt}
\affiliation{Argonne National Laboratory, Argonne, Illinois 60439, USA}
\affiliation{Department of Physics and Astronomny, Northwestern University, Evanston, Illinois 60208, USA}
\author {A. Rudenko}
\affiliation{J.R. Macdonald Laboratory, Department of Physics, Kansas State University, Manhattan, Kansas, 66506, USA}
\author{A. Marinelli}
\affiliation{SLAC National Accelerator Laboratory, Menlo Park, California 94025, USA}
\author{D. Moonshiram}
\affiliation{Argonne National Laboratory, Argonne, Illinois 60439, USA}
\author{T. Osipov}
\affiliation{SLAC National Accelerator Laboratory, Menlo Park, California 94025, USA}
\author{D. Rolles}
\affiliation{J.R. Macdonald Laboratory, Department of Physics, Kansas State University, Manhattan, Kansas, 66506, USA}
\affiliation{Deutsches Elektronen-Synchrotron (DESY), 22607 Hamburg, Germany}
\author{N. Berrah}
\affiliation{Department of Physics, University of Connecticut, Storrs, Connecticut, 06269, USA}
\author{C. Bomme}
\affiliation{Deutsches Elektronen-Synchrotron (DESY), 22607 Hamburg, Germany}
\author{M. Bucher}
\affiliation{Argonne National Laboratory, Argonne, Illinois 60439, USA}
\affiliation{SLAC National Accelerator Laboratory, Menlo Park, California 94025, USA}
\author{G. Doumy}
\affiliation{Argonne National Laboratory, Argonne, Illinois 60439, USA}
\author{B. Erk}
\affiliation{Deutsches Elektronen-Synchrotron (DESY), 22607 Hamburg, Germany}
\author{K. R. Ferguson}
\affiliation{SLAC National Accelerator Laboratory, Menlo Park, California 94025, USA}
\author{T. Gorkhover}
\affiliation{SLAC National Accelerator Laboratory, Menlo Park, California 94025, USA}
\author{P. J. Ho}
\affiliation{Argonne National Laboratory, Argonne, Illinois 60439, USA}
\author{E. P. Kanter}
\affiliation{Argonne National Laboratory, Argonne, Illinois 60439, USA}
\author{B. Kr\"assig}
\affiliation{Argonne National Laboratory, Argonne, Illinois 60439, USA}
\author{J. Krzywinski}
\affiliation{SLAC National Accelerator Laboratory, Menlo Park, California 94025, USA}
\author{A. A. Lutman}
\affiliation{SLAC National Accelerator Laboratory, Menlo Park, California 94025, USA}
\author{A. M. March}
\affiliation{Argonne National Laboratory, Argonne, Illinois 60439, USA}
\author{D. Ray}
\affiliation{SLAC National Accelerator Laboratory, Menlo Park, California 94025, USA}
\author{L. Young}
\affiliation{Argonne National Laboratory, Argonne, Illinois 60439, USA}
\affiliation{The James Franck Institute and Department of Physics, The University of Chicago, Chicago, Illinois 60637, USA}
\author{S. T. Pratt}
\affiliation{Argonne National Laboratory, Argonne, Illinois 60439, USA}
\author{S. H. Southworth}
\affiliation{Argonne National Laboratory, Argonne, Illinois 60439, USA}
\
 \date{\today}

\begin{abstract}

The capability of generating two intense, femtosecond x-ray pulses with controlled time delay opens the possibility of performing time-resolved experiments for x-ray induced phenomena.  We have applied this capability to study the photoinduced dynamics in diatomic molecules. In molecules composed of low-Z elements, \textit{K}-shell ionization creates a core-hole state in which the main decay mode is an Auger process involving two electrons in the valence shell. After Auger decay, the nuclear wavepackets of the transient two-valence-hole states continue evolving on the femtosecond timescale, leading either to separated atomic ions or long-lived quasi-bound states. By using an x-ray pump and an x-ray probe pulse tuned above the \textit{K}-shell ionization threshold of the nitrogen molecule, we are able to observe ion dissociation in progress by measuring the time-dependent kinetic energy releases of different breakup channels.  We simulated the measurements on N$_2$ with a molecular dynamics model that accounts for \textit{K}-shell ionization, Auger decay, and the time evolution of the nuclear wavepackets.  In addition to explaining the time-dependent feature in the measured kinetic energy release distributions from the dissociative states, the simulation also reveals the contributions of quasi-bound states.
\end{abstract}

\maketitle

\section{Introduction}

Newly developed capabilities at x-ray free-electron lasers (XFELs) allow the production of two intense x-ray pulses with controlled time delay \cite{Lutman2013,DeNinno2013,Allaria2013,Petrillo2013,Hara2013,Marinelli2015}. Those capabilities open the possibility to track, with femtosecond resolution, x-ray induced phenomena in matter. 
\begin{figure}[h]
\begin{center}
\includegraphics[trim=6cm 7cm 6cm 2cm,clip,width=8.5cm]{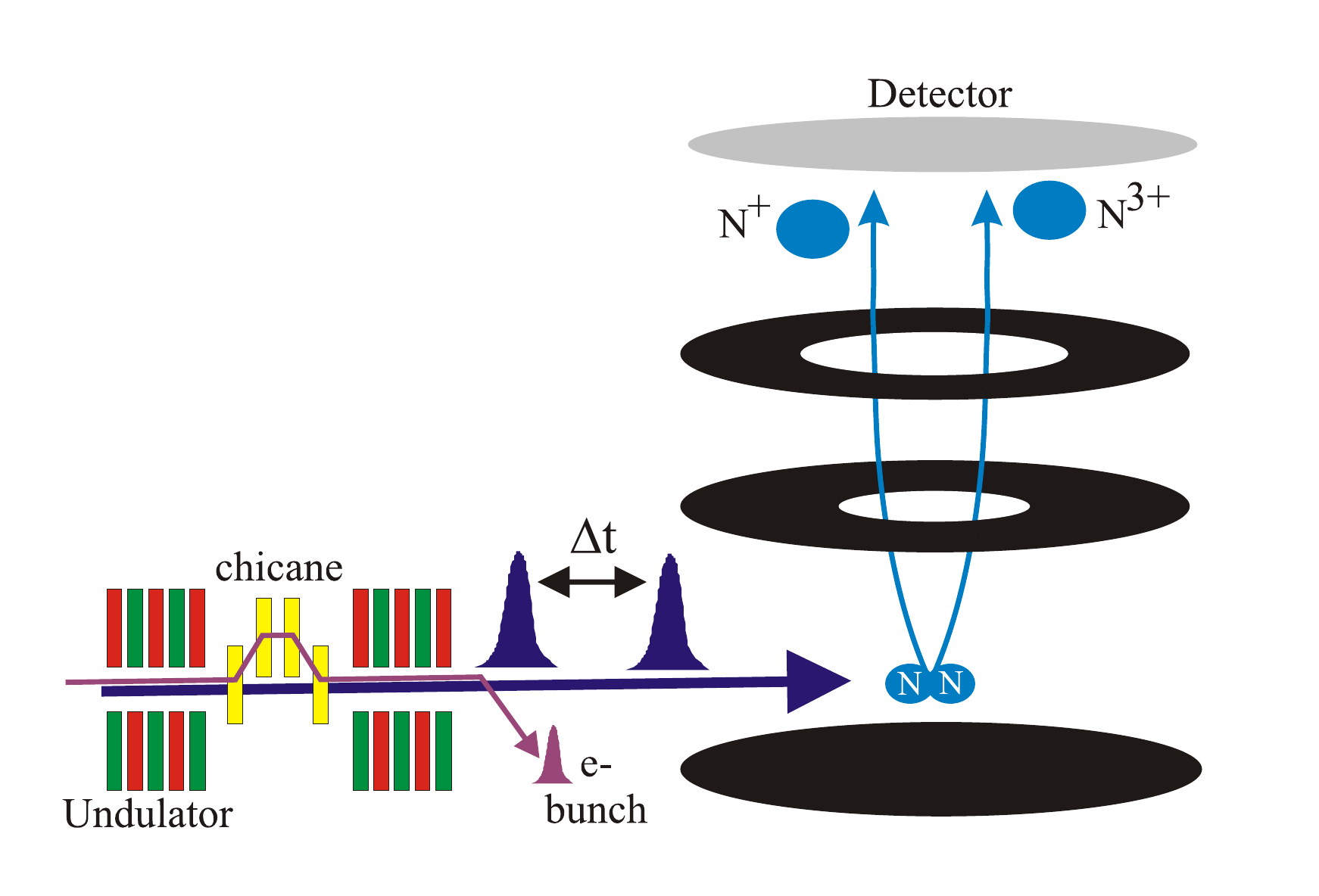}
\caption{Scheme of the experimental setup. Two 10-fs x-ray pulses are produced with a controlled time delay, as described in Ref. \cite{Lutman2013}. The pump and probe x-ray energies were both $\sim$700 eV, i.e., far above the 410-eV \textit{K}-shell ionization energy of N$_2$.  Three time delays are considered: 4, 29, and 54 fs. The first pulse triggers \textit{K}-shell ionization and Auger decay of N$_2$, dominantly producing the N$^{+}$-N$^{+}$ breakup channel. \textit{K}-shell ionization and Auger decay of one of the N$^+$ ions by the probe pulse converts it to N$^{3+}$. The ions are recorded in coincidence with a position sensitive detector and the kinetic energy release (KER) is determined. The time-dependent pump-probe signal is observed in the measured KER of the N$^{+}$-N$^{3+}$ breakup channel.}
\label{fig:scheme}
\end{center}
\end{figure}

In recent years, tremendous progress has been made in the study of molecular nuclear dynamics induced by strong-field ionization \cite{Ergler2005,De2010,Bocharova2011} and valence photoabsorption \cite{Gagnon2007,Sandhu2008,Kelkensberg2009,Sansone2010}. These studies consist of a wide variety of time-resolved experiments, in which the pump or/and the probe is a strong-field optical laser. However, extracting information from the experimental schemes involving strong-field interactions requires theoretical modeling to account for the induced strong-field effects \cite{Magrakvelidze2012}. With the advent of FELs, a new generation of pump-probe schemes was conceived that uses two XUV pulses \cite{Rudenko2010,Jiang2010,Jiang2013,Schnorr2014}. The high flux of photons in a femtosecond pulse generated by FELs is crucial to obtain the needed signal in the pump-probe scheme, in which absorption of at least one photon from each pulse by the same molecule is required. By controlling the photon energy and the time delay between the XUV pulses, it is possible in principle to excite and select a particular state in the molecule and track the nuclear dynamics with the probe pulse, with the advantage of avoiding strong-field effects or multi-photon excitations. This opens the possibility of studying nonadiabatic effects at conical intersections \cite{kowalewski2015}, imaging or clocking nuclear wave packets \cite{Picon2011,GonzalezCastrillo2012}, and exploring isomerization processes \cite{Jiang2010_2,Jiang2013,Rescigno2012}. Moreover, the generation of two XUV pulses with table-top sources is also possible, by using high-order high-harmonic generation (HHG) \cite{Kling2008}.  Great progress has been made in performing ultrafast measurements with HHG sources \cite{Tzallas2011,Fabris2015}, despite the challenge of generating enough photon flux in the inefficient nonlinear conversion process.

Schemes for producing two intense x-ray pulses with controlled time delay at XFELs have been developed only in recent years, and very few experiments have been performed on molecules using two x-ray pulses for ultrafast measurements \cite{Chelsea2015,Picon2016}. Those experiments aim at providing novel information on x-ray-induced processes by using an x-ray probe pulse. Selecting the x-ray energies of the two pulses enables site-selective excitation and probing of the evolving system, i.e., by following changes in the local electronic configuration and nuclear geometry at a particular site of the molecule \cite{Picon2016}. 

One of the applications for such capabilities is the understanding of molecular fragmentation induced by x-rays, a process of fundamental interest and a relevant factor for the radiation damage induced during the structural imaging of biomolecules at XFELs \cite{Neutze2000,Lomb2011,Nass2015}.  Molecular ionization and fragmentation follow absorption of a single synchrotron x ray \cite{Nenner1996,Dunford2012,Guillemin2015} or absorption of multiple x rays using XFEL radiation \cite{Hoener2010,Fang2012,Osipov2013,Erk2013,Erk2014}.  In the present experiment, we exploited the femtosecond time resolution of XFEL pulses, but the fluences of the pump and probe pulses were designed to observe single-photon pump and single-photon probe events.  { We applied this scheme to observe x-ray-induced molecular fragmentation.}

Basic information on the dynamics involved in molecular fragmentation by x rays can be obtained by measurements and calculations on small molecules. Here we demonstrate an x-ray/x-ray pump-probe scheme to follow the x-ray induced molecular fragmentation of diatomic molecules by using a coincidence ion momentum imaging spectrometer \cite{Ullrich2003}. This scheme, together with theoretical calculations, provides access to the intermediate quantum nuclear wavepackets propagating during and after electronic relaxations.

In the general case, absorption of an x ray by the molecule creates a core-hole state by the promotion of an inner-shell electron into a valence-like state or the continuum. The core-hole state is highly unstable and decays rapidly, with a lifetime on the order of hundreds of attoseconds to a few femtoseconds. The decay occurs either by nonradiative processes, such as Auger decay, or by radiative processes, such as x-ray fluorescence. The decay of the core-hole may induce other inner-shell or valence-shell holes in higher energy shells, consequently triggering other nonradiative or radiative processes. This picture is far more complicated in molecules than in atoms, because the nuclear motion must also be considered. The removal of electrons during nonradiative processes may produce highly-charged states with strong dissociative behavior. In some cases, the nuclear dynamics cannot be neglected even during the characteristic core-hole lifetimes. 

Here we implement a pump-probe scheme to study the fragmentation of the nitrogen molecule upon the absorption of an x-ray photon. In molecules composed of low-Z elements, such as N$_2$, the ionization of a \textit{K}-shell electron creates a core-hole state in which the main decay mode is an Auger process involving two electrons in the valence shell. Therefore, after Auger decay, excited states with two holes in the valence shell are populated. The core-hole lifetime in the nitrogen molecule is about 3.6 fs. With the present experimental setup, the time resolution { of $\sim$10 fs} is not sufficient to measure dynamical effects during the Auger decay. However, it is possible to follow the induced nuclear motion that proceeds after Auger decay. 

By measuring kinetic energy releases (KERs) at different time delays between the pump and the probe, we are able to resolve the nuclear dynamics for a specific breakup channel. {We previously reported on an x-ray/x-ray pump-probe experiment on XeF$_2$ molecules \cite{Picon2016}.  Due to the complexity of the vacancy cascade, a classical model based on Coulomb repulsion was used to model breakup of the molecular ions.  For the case of N$_2$ studied here, we present a quantum mechanical treatment of \textit{K}-shell photoionization by the x-ray pump, Auger decay, propagation of nuclear wavepackets on the N$_2^{2+}$ potential curves, and the time-dependent KERs induced by the x-ray probe.}  Our simulation is in agreement with the measured KER and allows an interpretation of the underlying mechanism.

{With our present time resolution and ion coincidence spectrometer we are able to observe breakup of N$_2^{2+}$ into atomic ions and explain our measurements with a quantum mechanical simulation.  Our results are encouraging for future experiments at XFELs that will generate few-femtosecond x-ray pulses at high repetition rates \cite{LCLS2}.  These advances will allow coincidence detection of electrons with ions and enable the possibility to select intermediate states, explore electronic relaxation processes during the Auger decay, and observe nuclear nonadiabatic couplings.}

\section{Experiment}
The pump-probe setup is illustrated in Fig. \ref{fig:scheme}. The experiment was performed at the Linac Coherent Light Source (LCLS) using the Atomic, Molecular, and Optical (AMO) physics end station \cite{Ferguson2015}.  The pump and probe x-ray pulses with controlled time delay were generated { at 120 Hz} using single electron bunches and split undulator sections as described in Ref. \cite{Lutman2013}. We used two 10-fs x-ray pulses with energies $\sim$700 eV, that is, far above the $\sim$410-eV \textit{K}-shell ionization energy of N$_2$ \cite{Chen1989} and of the \textit{K}-shell ionization energies of atomic nitrogen ions \cite{Garcia2009}.  The two pulse energies were roughly balanced, and the combined pulse energies were $\sim$33 $\mu$J.  The x-ray transport optics have an efficiency $\sim$20\%, and the focal spot area was $\sim$5 $\mu$m$^{2}$.  The combined peak intensity for the overlapping pump and probe pulses was $\sim$1.3 $\times$ 10$^{16}$ W/cm$^{2}$.  Measurements were made with pump-probe delays of 4, 29, and 54 fs.  { While higher time resolution and additional time points are desirable, the available beamtime and need to record ion-ion coincidences with sufficient counting statistics were limitations.}

\begin{figure}[h]
\begin{center}
\includegraphics[width=8.5cm]{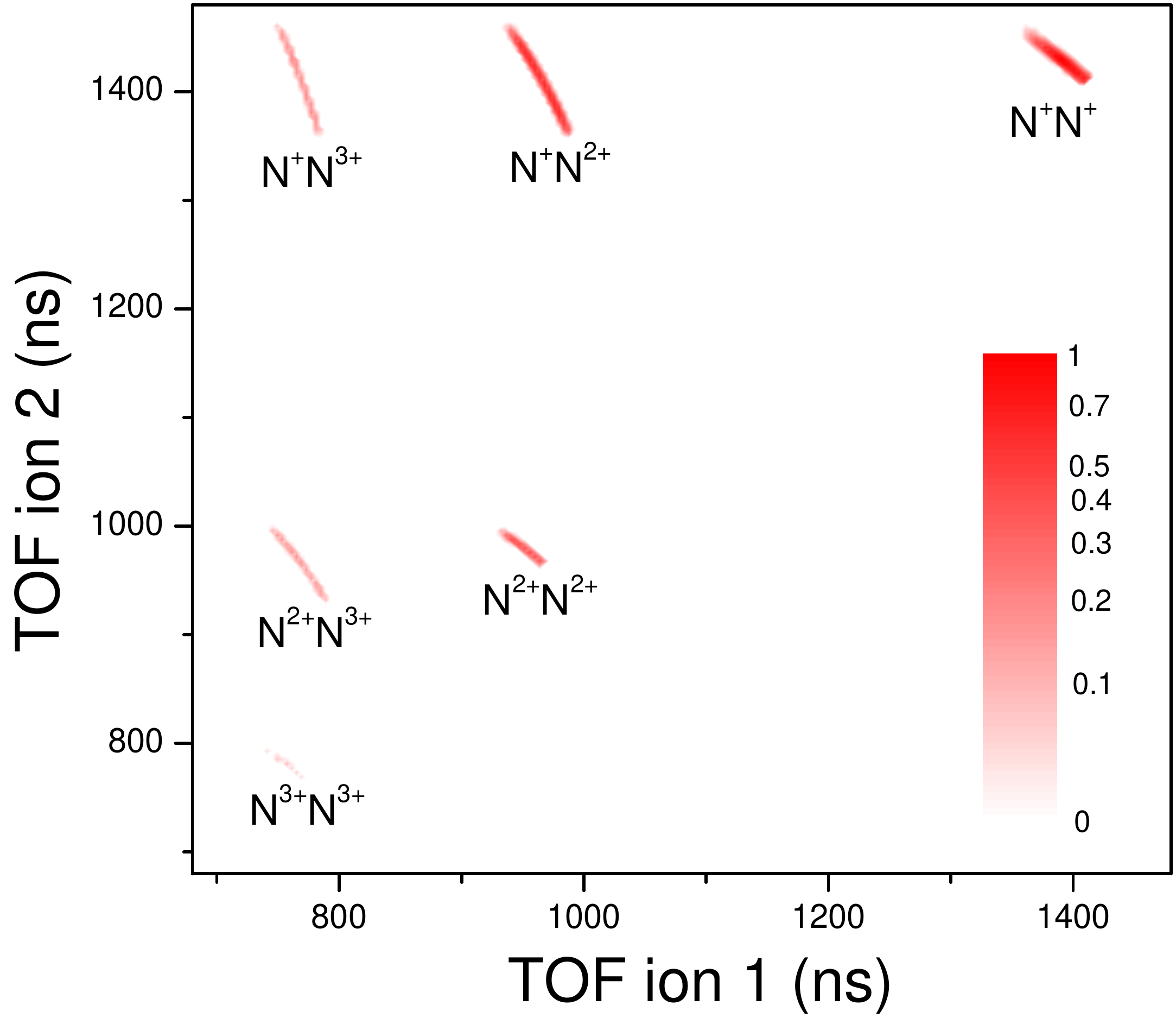}
\caption{Photoion-photoion coincidence map showing the ion breakup channels of N$_2$ from x-ray-pump/x-ray-probe measurements.  The axes are the times-of-flight (TOFs) of the two ions that are detected in coincidence and a momentum conservation filter has been applied.  The color intensity scale is normalized to 6720 maximum counts and the total counts is 49680.}
\label{fig:pipico}
\end{center}
\end{figure}

The ions created by x-ray fragmentation of N$_2$ molecules were projected onto a hexanode delay-line detector by a homogeneous electric field and detected in coincidence. Their momenta were determined from their times-of-flight and positions on the detector \cite{Ullrich2003}.  Figure \ref{fig:pipico} shows the N$^{q_1^+}$-N$^{q_2^+}$ breakup channels in a photoion-photoion coincidence map.  The kinetic energy release (KER) distributions of the breakup channels are sensitive to the charge states and ion-ion separations, so the KERs of the pump-probe events depend on the time delay and yield information on the system during the dissociation process.  The main outcome of our measurements and data analysis is observation of the delay-dependent population transfer from the N$^{+}$-N$^{+}$ channel produced by x-ray absorption within the pump pulse to the N$^{+}$-N$^{3+}$ final state resulting from subsequent probe-pulse x-ray absorption by one of the N$^{+}$ ions.  { As explained below, the KERs also contain contributions from quasi-bound N$_2^{2+}$ states.}



\section{Results}
The potential energy curves (PECs) of the two-hole states of N$_{2}^{2+}$ have been characterized before, and some states are dissociative while others are quasi-bound \cite{Pandey2014}.  Some quasi-bound states are long-lived and the dication molecule remains undissociated when detected. This is well known in ion time-of-flight spectra of N$_2$ that show a narrow peak from zero-momentum N$_{2}^{2+}$ ions and broad wings from energetic N$^{+}$ ions \cite{Fang2012,Pandey2014}.  The dissociative states are correlated with either two separated N$^{+}$ ions or separated neutral N atom and N$^{2+}$ ion. The Auger decay in N$_2$ has been studied through Auger-electron \cite{Moddeman1971,Eberhardt1987}, KER \cite{Saito1987,Lundqvist1996}, and ion-electron coincidence spectroscopy \cite{Weber2001,Semenov2010,Rolles2010}. For the dissociative states in the N$^{+}$-N$^{+}$ breakup channel, the range of the KER distribution reflects the energy difference between the final two-hole states after Auger decay and the separated ion products. In the past, our knowledge about the transient states after x-ray absorption has mainly relied on measurements of final products (electrons, ions, or photons) and the proper theoretical modeling. In the present work, a second x-ray pulse is used to gain insight on the transient states in real time with $\sim$10 fs resolution. 

By absorbing a second x-ray from the probe pulse, a second 1s electron from one atomic site is removed, and that predominantly triggers a second Auger process. The probe-induced \textit{K}-shell ionization and Auger decay predominantly increases the charge state of the system by 2. In the following, we show theoretical simulations that demonstrate how the probe pulse maps the dication nuclear wavepackets onto the KER of the tetracation states. Hence, by changing the time delay and measuring the KER, we can retrieve nuclear motion information of the low-charge-state channels. 



KER spectra of molecular nitrogen after absorption of a single x ray have been reported \cite{Weber2001,Semenov2010}; the main detected breakup channels are N$^{+}$-N$^{+}$ and N$^{+}$-N$^{2+}$. Production of the N$^{+}$-N$^{2+}$ channel is due to many-electron correlation effects associated with shake-off processes in the Auger decay step \cite{Eberhardt1987,Saito1987}. In our experiment, due to the absorption of two photons \cite{Fang2012}, we measure higher charge states; N$^{+}$-N$^{3+}$, N$^{2+}$-N$^{2+}$, N$^{2+}$-N$^{3+}$, and N$^{3+}$-N$^{3+}$ (see Fig. \ref{fig:pipico}). Those charge states can be produced by the absorption of two photons either in the pump or in the probe pulse, but also by pump-probe events, i.e. the absorption of one photon from the pump pulse and the absorption of a second photon from the probe pulse. 

\begin{figure}[h]
\begin{center}
\includegraphics[width=9cm]{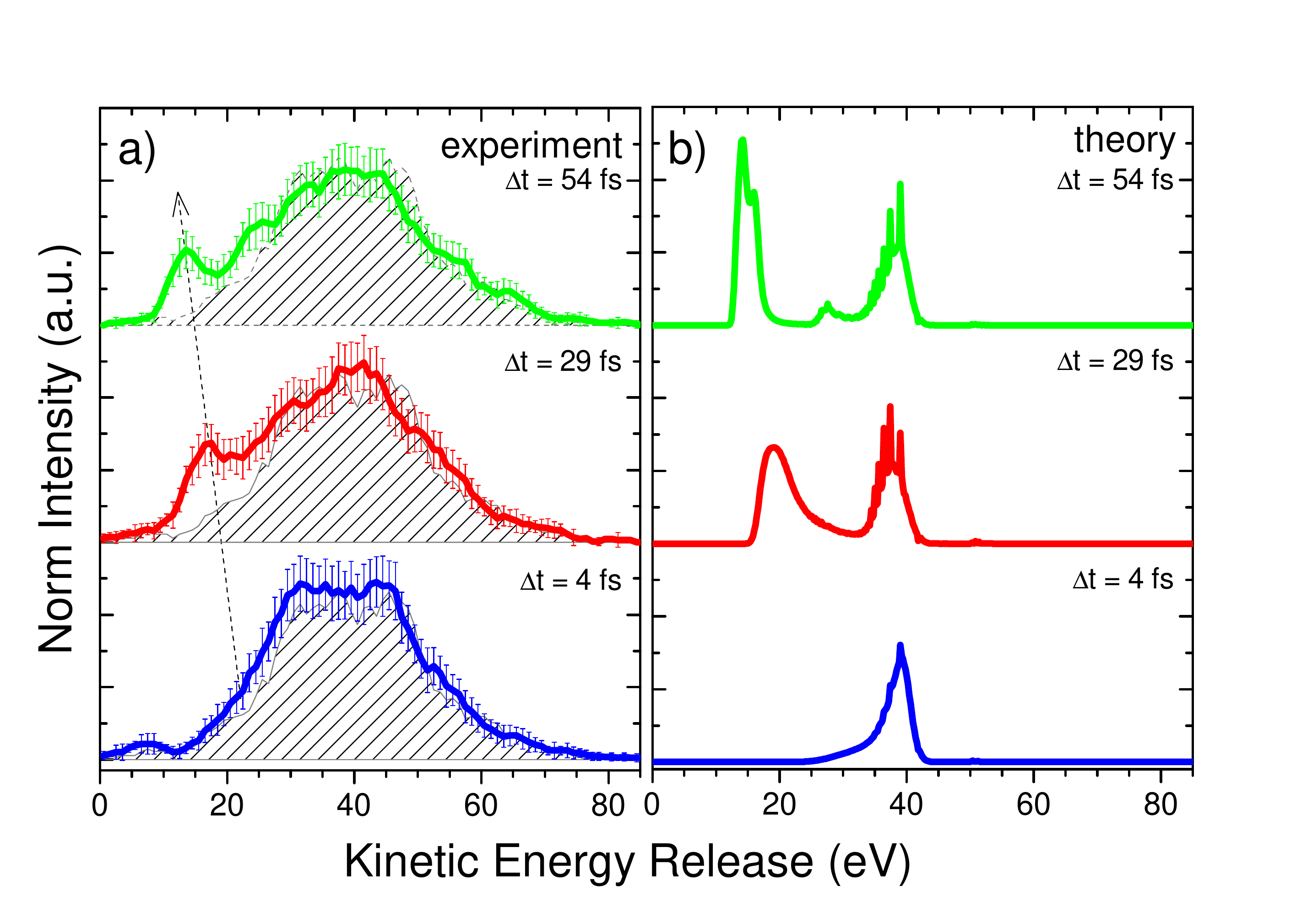}
\caption{Kinetic energy release for the N$^{+}$--N$^{3+}$ breakup channel at three time delays: 4 fs (blue), 29 fs (red), and 54 fs (green). (a) Experimental measurements. The shaded areas are the pump-only measurements.  The arrow indicates the peak due to pump-probe events from dissociative intermediate states. (b) Theoretical calculations for pump-probe events.  The lower-energy peaks in the 29 fs and 54 fs calculations are from N$^{+}$--N$^{+}$ dissociative intermediate states, and the peaks at $\sim$38 eV are from N$_2^{2+}$ quasi-bound states.  The dissociative and quasi-bound states are unresolved in the 4 fs simulation.}
\label{fig:KER}
\end{center}
\end{figure}

N$^{+}$-N$^{+}$ is the dominant channel after pump excitation, therefore we expect pump-probe signals in both the N$^{2+}$-N$^{2+}$ and the N$^{+}$-N$^{3+}$ channels. At short internuclear distances, the valence electrons are delocalized and interatomic interactions are strong. There, the probe-induced Auger decay populates either the N$^{2+}$--N$^{2+}$ or the N$^{+}$--N$^{3+}$ channel. As the separation between the ions increases, however, the probe interacts with an isolated N$^{+}$ ion and enhances the N$^{+}$--N$^{3+}$ channel. We show in Fig. \ref{fig:KER}(a) the KER for the N$^{+}$--N$^{3+}$ breakup channel at three time delays. The main structure is centered near 40 eV. This structure could come from two-photon absorption or from single-photon absorption, since synchrotron x-ray measurements show that N$^{3+}$ ions are produced at photon energies much higher than the \textit{K}-shell ionization threshold \cite{Stolte1998}. The relative yield of N$^{3+}$ ions obtained in our experiment is in the same order as the high-energy synchrotron data of Ref. \cite{Stolte1998}. However, in Fig. \ref{fig:KER}(a), we also clearly observe a peak that shifts towards lower KER as the time delay increases. Our proposed explanation is that, at shorter time delays, such as our 4 fs measurement, the dissociative states of N$^{+}$-N$^{+}$ are still close to the internuclear distance of N$_2$. Then, the interaction with the probe promotes those states into the N$^{+}$-N$^{3+}$ breakup channel with a high Coulomb repulsion force and high KER. However, at longer time delays, the internuclear distances of the dissociative states have increased, and when promoted to the N$^{+}$-N$^{3+}$ channel, the Coulomb repulsion is not as strong. That pathway ends up with low KER, in accord with the feature observed in Fig. \ref{fig:KER}(a). In order to support this interpretation and gain insight of the underlying mechanism, we have performed quantum-mechanical simulations that are described in the following. 


\section{Theory and simulations}
\begin{figure}[h]
\begin{center}
\includegraphics[width=8.cm]{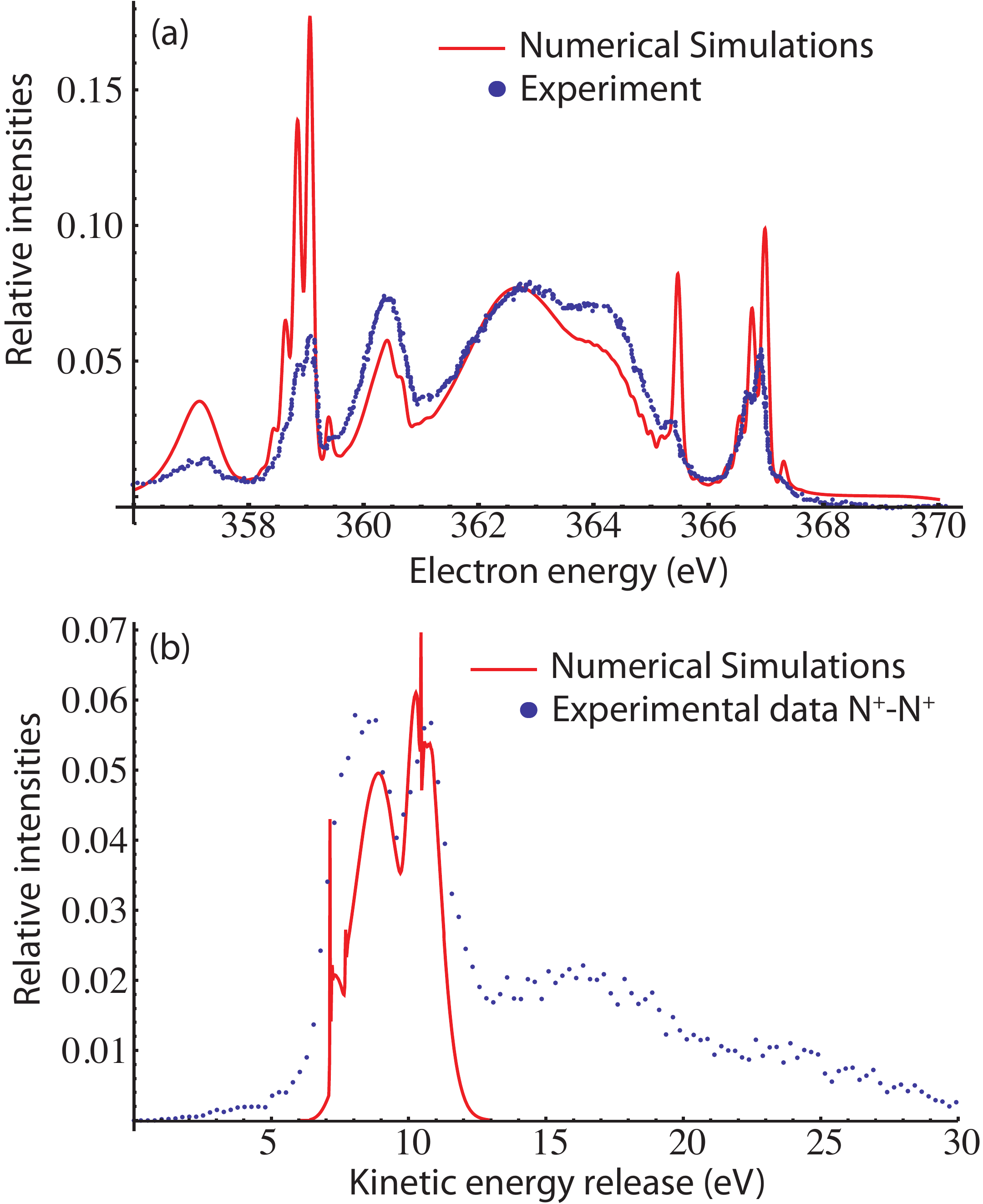}
\caption{(a) Auger electron spectrum after \textit{K}-shell ionization of N$_2$. Blue dot points, experimental data extracted from Ref. \cite{Sorensen2008}. Red line, results from theoretical calculations presented in this paper. (b) Kinetic energy release of the N$^{+}$-N$^{+}$ breakup channel. Blue dot points and red line are experimental data and theoretical results presented in this paper, respectively. No time dependence was observed in this channel.}
\label{fig:NN_sim}
\end{center}
\end{figure}

Our time-resolved measurements can be theoretically described by a molecular dynamics model that accounts for the N$^{+}$-N$^{+}$ channel. We need to account for both the \textit{K}-shell ionization induced by the pump pulse and the Auger decay that populates a manifold of final dication states, concurrently with the time evolution of the nuclear wavepackets. We use Fourier-transform limited pulses with FWHM of 10 fs for the theoretical simulation. In the experiment we use Self-Amplified-Spontaneous-Emission (SASE) pulses, but the bandwidths of the pulses are not relevant in this case, because the x-ray energies are far above the \textit{K}-shell ionization threshold, so only core-hole states with a high-energy photoelectron are produced. The electric dipole moments needed in the ionization are calculated with Cowan's Hartree-Fock program \cite{LosAlamos1,LosAlamos2}. The Auger transitions from the core-excited state to the final dication states are taken from Ref. \cite{Agren1981}. The time evolution of the system is calculated by solving the equations of motion for amplitudes (EOM). EOM are obtained by projecting the whole Hilbert space onto the electronic levels in which the main dynamics is confined. In particular, we expand the wavefunction of the system as 
{
\begin{eqnarray}
&& \Psi({\bf X},{\bf R},t) = b_{i} ({\bf R},t)\Phi_{i}({\bf X},{\bf R}) + \nonumber \\
&& \int \!\! dE_{ph} b_{ce} (E_{ph},{\bf R},t)\Phi_{ce,E_{ph}}({\bf X},{\bf R}) + \nonumber \\
&& \sum_n\int \!\! \int \!\! dE_{ph} dE_a \, b_n(E_{ph},E_a,{\bf R},t)\Phi_{n,E_{ph},E_a}({\bf X},{\bf R}),\
\end{eqnarray}
where ${\bf X}$ stands for the electronic coordinates and ${\bf R}$ for the nuclear coordinates. $\Phi_{i}({\bf X},{\bf R})$ is the ground state, $\Phi_{ce,E_{ph}}({\bf X},{\bf R})$ is the core-excited state, and $\Phi_{n,E_{ph},E_a}({\bf X},{\bf R})$ are the final dication states. $E_{ph}$ labels a particular quantum state with a specific photoelectron energy, while $E_{a}$ refers to a specific Auger electron energy.} For the nitrogen molecule, ${\bf R}$ is reduced to one dimension and from here on we take this variable as the internuclear distance between the nitrogen atoms. Hence, the amplitudes $b$ depend on the internuclear distance and are nuclear wavepackets propagating in the corresponding electronic energy curves. Nonadiabatic effects have not been taken into account. The energy curves for the ground state, core-excited state, and the dication states have been obtained from Refs. \cite{Herzberg1965,Ehara2006,Pandey2014}. In our simulations, nine low-lying excited states of the dication molecule are included, with symmetries $^1\Sigma_g^+$, $^3\Sigma_u^+$, $^1\Pi_u$, $^3\Pi_g$, $^1\Sigma_g^+$, $^1\Delta_g$, $^1\Pi_g$, $^1\Sigma_u^+$, and $^1\Sigma_g^+$ \cite{Pandey2014}. Those excited states present the largest Auger decay rates after \textit{K}-shell ionization \cite{Agren1981}. As seen in Fig. \ref{fig:NN_sim}(a), the numerical results reproduce the line shape of the Auger spectrum measured in a previous experiment \cite{Sorensen2008}.  { Note that the quality of the calculated Auger energies mainly depends on the potential energy surfaces that are used, while the ratio of the peaks for different electronic levels mainly depend on the Auger decay rate calculations.} Figure \ref{fig:NN_sim}(b) shows the calculated KER for the N$^{+}$-N$^{+}$ channel compared with our experimental data. The two main peaks centered at 10 eV are well described. The high ($>$13 eV) KER of the spectrum originates from high-lying excited states of the dication \cite{Pandey2014} that are not included in the simulations. The high KER part is very difficult to model because of the numerous excited states and Auger transitions that need to be included. However, the high KER part is not needed to qualitatively explain the time-dependent feature observed in Fig. \ref{fig:KER}(a), as we will argue in the following.

The theoretical simulations describe the essential observables due to the absorption of the pump pulse. In order to account for the effects of the probe pulse we make the following assumptions: 1) the probe pulse increases the charge state of the system by two, and 2) the nuclear wavepacket is promoted into a potential energy curve that is mainly dominated by the Coulomb repulsion between N$^+$ and N$^{3+}$. The first assumption is not exact, because other processes, such as valence ionization, x-ray fluorescence, and shake off, might result in charge-state changes other than two.  However, in nitrogen, those processes are at least one order of magnitude less frequent than the normal Auger decay. In the second assumption, we consider only the case in which the Auger decay is mainly localized in one site and there is no charge redistribution. By using the classical model, described in Ref. \cite{Ryufuku1980}, we can estimate the internuclear distance at which this assumption is valid, i.e. when the charge redistribution from one site to the other site is forbidden. Using that model we obtain an internuclear distance of 6.3 a.u.. By a classical break-up of two N$^+$ ions starting at the equilibrium distance, we estimate that an internuclear distance of 6.3 a.u. is reached around 20 fs. Therefore, the second assumption for the shortest time delay at 4 fs is not necessarily valid, but it should be a good approximation for dissociative states at the 29 and 54 fs time delays. The KER of the N$^+$-N$^{3+}$ channel is then directly obtained from the nuclear wavepacket of the N$^+$-N$^{+}$ channel. We map the internuclear distance distribution, given by the nuclear wavepacket of N$^+$-N$^{+}$ in a specific time delay, into the KER of the N$^+$-N$^{3+}$ channel (a similar model was used for calculating KERs in Ref. \cite{Magrakvelidze2012}). We show the numerical results for the KER of N$^+$-N$^{3+}$ in Fig. \ref{fig:KER}(b) and observe the time-dependent peak moving towards lower KER with increasing delay, as observed in the experiment. From our simulations, we can assign the origin of that peak to the dissociative states after the Auger decay process. At 4 fs, contributions from both quasi-bound and dissociative states are located at large KER, around 38 eV. However, for longer time delays, the dissociative states move out of the equilibrium distance, which is reflected in the time-dependent peak moving towards the low KER distribution.  The multipeak structure remaining around 38 eV arises from quasi-bound states populated after the Auger decay. Some of them are very long-lived and impossible to observe in the measured KER of N$^+$-N$^{+}$ (they arrive at the detector without dissociating, as a single N$_2^{2+}$ ion). However, the probe pulse can promote the quasi-bound states into the N$^+$-N$^{3+}$ dissociative channel and they may be detectable. Unfortunately, in our experiment, we did not have enough statistics to resolve this multipeak structure, because it lies on the top of the main structure that is most likely dominated by one-photon absorption. For very long time delays, we expect the N$^+$-N$^{+}$ KER distribution, see Fig. \ref{fig:NN_sim}(b), to mainly be mapped into the N$^+$-N$^{3+}$ channel. The high KER range of the N$^+$-N$^{+}$ channel that has not been modeled should then not interfere with the observed time-dependent peak, since it produces a broad distribution in the higher part of the transient KER spectrum.

We should remark that if the pump step involves low-energy continuum or resonant core-hole states, the SASE bandwidth ($\sim$2-3 eV in the present experiment) will broaden the transient nuclear wavepackets.  That is not a factor in the present case of N$_2$ excited far above threshold.  Therefore, although the theoretical model is not complete, it clearly explains the observed peak moving towards lower KER in the N$^+$-N$^{3+}$ breakup channel. The theory allows the assignment of this peak to the dissociative, low-lying excited states of N$^+$-N$^{+}$. Moreover, the simulations also offer a broader perspective of the information we could extract in the general time-resolved experiment, such as the possibility to measure the quasi-bound excitation after molecular Auger decay or to do nuclear wavepacket imaging on the femtosecond timescale.  We also note that we performed the same experiment using $\sim$700-eV  pump and probe pulses on oxygen molecules that have a \textit{K}-shell ionization energy $\sim$543 eV.  We observe a similar time-dependent peak moving towards lower KER in the O$^+$-O$^{3+}$ breakup channel, supporting the underlying mechanism discussed above.


\section{Conclusions}
In conclusion, we demonstrate the feasibility of using two x-ray pulses for studying molecular nuclear dynamics triggered by the absorption of an x-ray photon. A first x-ray pulse is absorbed by the molecule and initiates the dynamics under study. Then a second x-ray pulse probes the molecular dynamics at variable time delays. The pump and probe pulses have similar photon energies tuned well-above the \textit{K}-shell ionization threshold. Here we use this scheme to study the molecular fragmentation of nitrogen molecules. The probe-induced Auger decay maps the time-dependent nuclear wavepacket in a particular charge-state channel onto the Coulombic repulsive potential curve corresponding to a higher-charge-state channel, resulting in a delay-dependent KER. Hence, by measuring KERs at different time delays, we are able to resolve femtosecond nuclear dynamics in a specific breakup channel. Quantum-mechanical simulations on nitrogen are in agreement with the measured KER and allow the interpretation of the underlying dynamics. This scheme fully exploits the recent capabilities at XFELs in order to generate two intense x-ray pulses with controlled time delay, and it is not affected by the temporal or the energy jitter of the pulses, ideal for SASE pulses. This work opens the possibility of designing time-resolved experiments with two x-ray pulses based on KER spectroscopy, suitable for studying x-ray-induced nuclear dynamics in diatomic molecules. 

Future capabilities with much shorter pulses will enable the possibility to explore the nuclear wavepacket propagation during the same time scale as Auger processes. This might open the possibility to explore the role of Auger processes in the coherent evolution of the nuclear wavepackets. Also, future x-ray sources with high-repetition rate will enable electron-ion coincidence measurements and, therefore, the retrieval of electronic configuration information. This will open the door of monitoring the nuclear wavepacket propagating via nonadiabatic crossings. { In particular, it will be very interesting to explore the non-adiabatic effects on triatomic molecules and extend our theoretical approach in order to consider not only the non-adiabatic couplings during nuclear propagation, but also more than one nuclear degree of freedom.}


\begin{acknowledgments}
A.P. and S.H.S acknowledge discussions with Masahiro Ehara and Kiyoshi Ueda about the core-excited potentials. This material is based upon work supported by the U.S. Department of Energy, Office of Science, Basic Energy Sciences, Chemical Sciences, Geosciences, and Biosciences Division and supported the Argonne group under contract no. DE-AC02-06CH11357, A.R. and D.R. under contract no. DE-FG02-86ER13491, and N.B. under contract no. DE-SC0012376. D.R. also acknowledges support from the Helmholtz Gemeinschaft through the Young Investigator Program. Use of the Linac Coherent Light Source (LCLS), SLAC National Accelerator Laboratory, is supported by the U.S. Department of Energy, Office of Science, Office of Basic Energy Sciences under Contract No. DE-AC02-76SF00515.
\end{acknowledgments}



\end{document}